\documentclass{icrc29}
\usepackage{graphicx,amssymb,amsmath,times,url}
\newcommand{\Xmax}{\ensuremath{X_\text{max}}}

\begin{document}
%Title of paper
\title[UHECR Composition Using HiRes-II]{UHECR Composition
  Measurements Using the HiRes-II Detector}
\author[D.R. Bergman, for the HiRes Collaboration] {D.R. Bergman$^a$,
  for the HiRes Collaboration\\
  (a) Rutgers - The State University of New Jersey, Department of
  Physics and Astronomy, Piscataway, New Jersey, USA}
\presenter{Presenter: D.R. Bergman (bergman@physics.rutgers.edu), \ 
  usa-bergman-D-abs2-he14-poster}

\maketitle

\begin{abstract}
  While stereo measurements of extensive air showers allow a more
  precise determination of the depth of shower maximum and hence the
  composition of UHECR's, monocular measurements allow one to go much
  lower in energy.  Since the composition of UHECR seems to constant
  throughout the HiRes stereo energy range but changing just below it,
  this is not a trivial lowering of the energy threshold.  We fit the
  observed \Xmax\ distribution to a combination of expected proton and
  iron \Xmax\ distributions, using two different interaction models,
  to determine the relative fraction of light and heavy components
  throughout the HiRes monocular energy range.  Using a two component
  fit allows both the mean \Xmax\ and the width of the \Xmax\ 
  distribution to contribute composition measurement and allows us to
  deal with the \Xmax\ acceptance bias caused by limited elevation
  coverage.  An updated analysis from a larger data set will be
  presented in Pune.
\end{abstract}

\section{Introduction}

Recent measurements of the UHECR composition by the HiRes
Prototype/MIA experiment\cite{hrmia} (HiRes/MIA) and by HiRes in
stereo mode\cite{hrstereo} (HiRes Stereo), seem to indicate that the
cosmic ray composition is changing from heavy to light below $10^{18}$
eV, and then remains light above $10^{18}$ eV.  This interpretation
depends on the fact that HiRes/MIA measures a large elongation rate,
larger than expected from an unchanging composition, while HiRes
Stereo measures an elongation rate consistent with an unchanging
composition.  The two measurements barely overlap in energy range, and
the change in the elongation rate is in just this overlapping range.
One would like to observe the low energy, changing composition
becoming constant at higher energies, all in one experiment.  To do so
with the HiRes detector one must go lower in energy, which also
requires that one look at monocular data; lower energy showers will
only be observed from one site, if at all.

Unfortunately, the limited elevation coverage of the HiRes-II detector
biases the \Xmax\ acceptance, with the bias increasing at lower
energies.  The bias stems from the requirement that one find \Xmax\ 
within that extent of the shower observed in the detector.  Events
that are closer to the detector are more likely to have \Xmax\ above
the visible range, and thus be cut.  Furthermore, because lower
energies can only be observed close to the detector, events at these
energies will have a larger acceptance bias than those at higher
energies.  Iron showers at a given energy, will be more affected than
proton showers.  This bias precludes performing an elongation rate
analysis at energies below $10^{18}$ eV.

\section{The Two Component Fitting Method}

Instead of performing an elongation rate analysis, we have chosen to
fit the \Xmax\ distribution to a combination of \Xmax\ distributions
from proton and from iron primaries.  The proton and iron \Xmax\ 
distributions are generated by Monte Carlo (MC) simulation.  The
\Xmax\ distribution, in a given energy bin, is stored in a histogram,
and the data histogram is then fit to a combination of the proton and
iron MC histograms to find the proportion of each.  The fit is
performed using the binned maximum likelihood technique as implemented
in the HBOOK\cite{hbook} routine HMCLNL.  The statistical uncertainty
of the fit is taken from the width of $-2\log L$ as fit to a quadratic
in the region about the minimum.

Once the proton fraction in an energy bin is determined, one must
correct for the fact that proton and iron showers have different
acceptances at a given energy.  This is done using the acceptances
calculated from the same MC samples used to make the \Xmax\ 
distributions for fitting the data.  The observed events at a given
energy are split up into the proton events and the iron events
according to the proton fraction.  Each of these samples is then
corrected for the acceptance, giving the number of proton events and
the iron events one would see with a fixed aperture and no \Xmax\ 
bias.  These corrected numbers of events are then used to calculate
the true proton fraction.

\section{Preliminary Results}

We generated three sets of Monte Carlo (MC) data using
Corsika/QGSJet01\cite{corsika,qgsjet}.  These three sets all use the
broken power law fit to the Fly's Eye\cite{fes} stereo spectrum as an
input spectrum.  The sets differ according to the composition, with
one being pure proton, one being pure iron, and one having the mixed
composition implied by the elongation rate measurements compared to
QGSJet expectations.  This last set is also the one used in the
HiRes-II monocular spectrum analysis.  The fitting procedure itself
only uses the pure composition sets; the mixed composition set is used
for cross checking.

\begin{figure}[h]
  \begin{minipage}[t]{0.49\textwidth}
    \includegraphics[width=\textwidth]{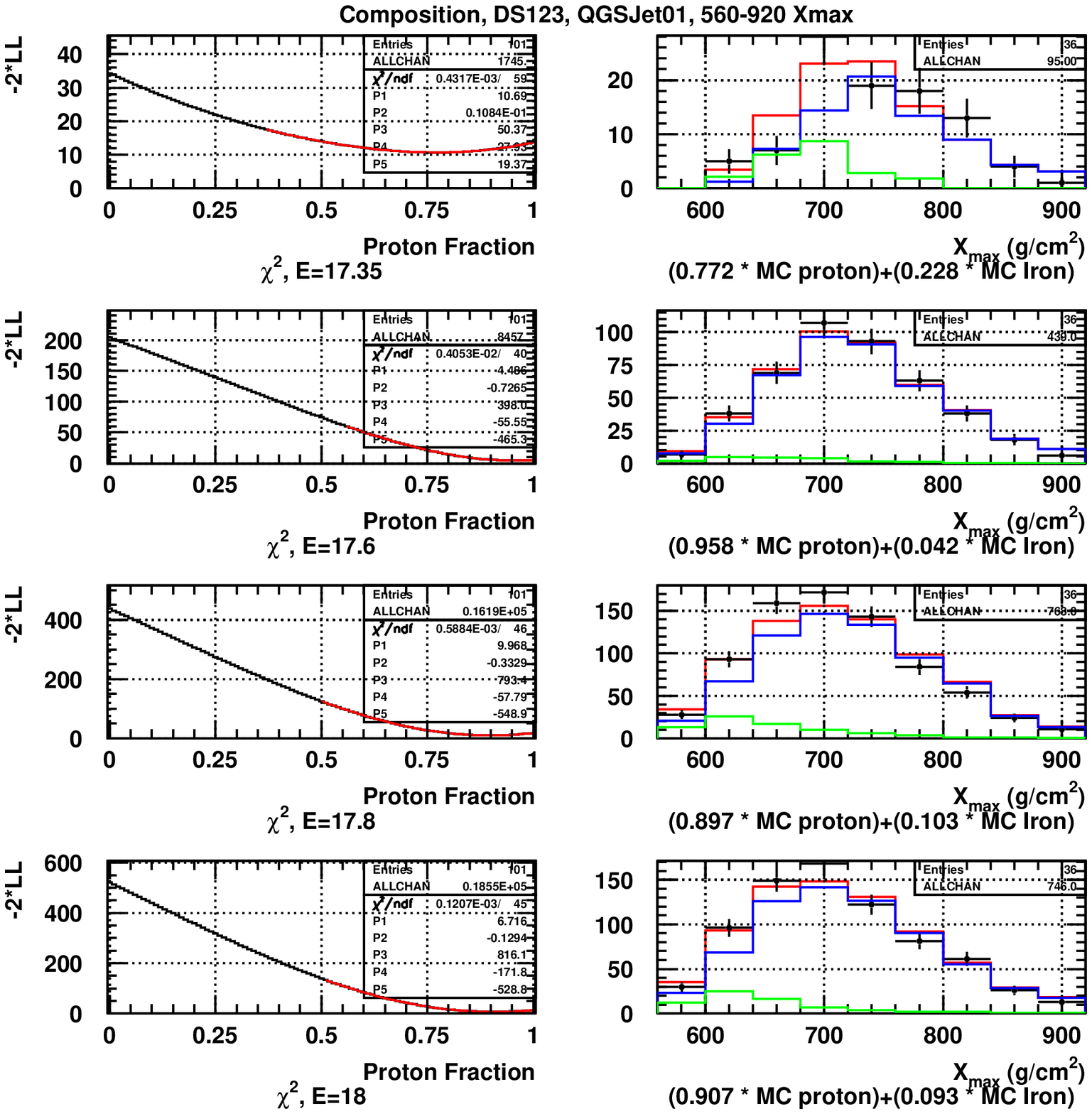}
  \end{minipage}\hfill%
  \begin{minipage}[t]{0.49\textwidth}
    \includegraphics[width=\textwidth]{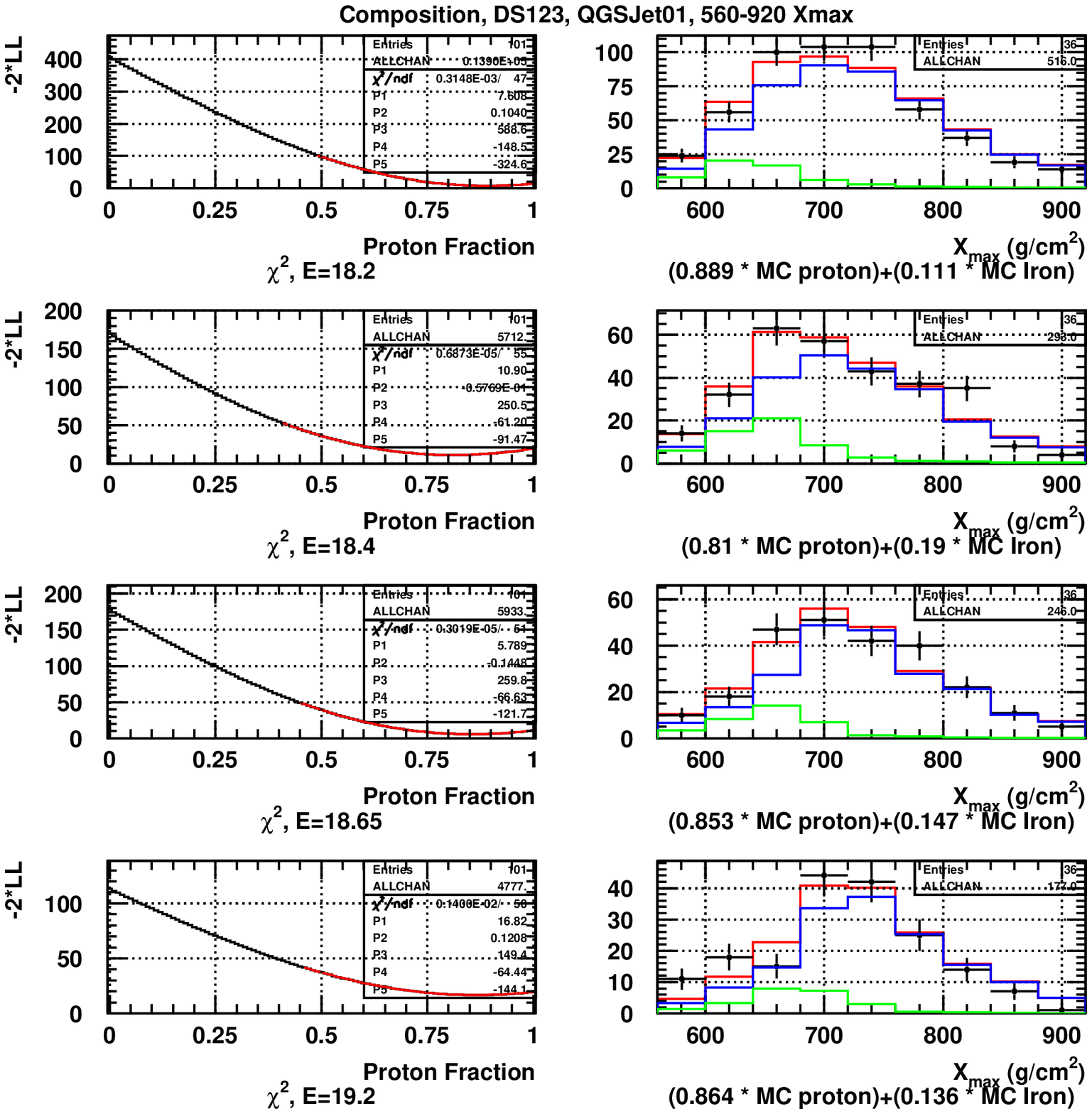}
  \end{minipage} 
  \caption{\label{fig:compfit} Two component composition fits of
    the \Xmax\ distribution in bins centered at $\log_{10}E = 17.35$,
    17.6, 17.8, 18, 18.2, 18.4, 18.65 and 19.2 ($E$ in eV).  The left
    side shows the quality-of-fit versus the proton fraction.  The
    right side shows the best fit (red) against the data (black
    points) along with the two components, protons (blue) and iron
    (green).}
\end{figure}

The results of the fits for proton fraction in the various energy
bands are shown in Figures~\ref{fig:compfit}.

\begin{figure}[h]
  \begin{center}
    \includegraphics*[width=0.75\textwidth,angle=0,clip]
                                    {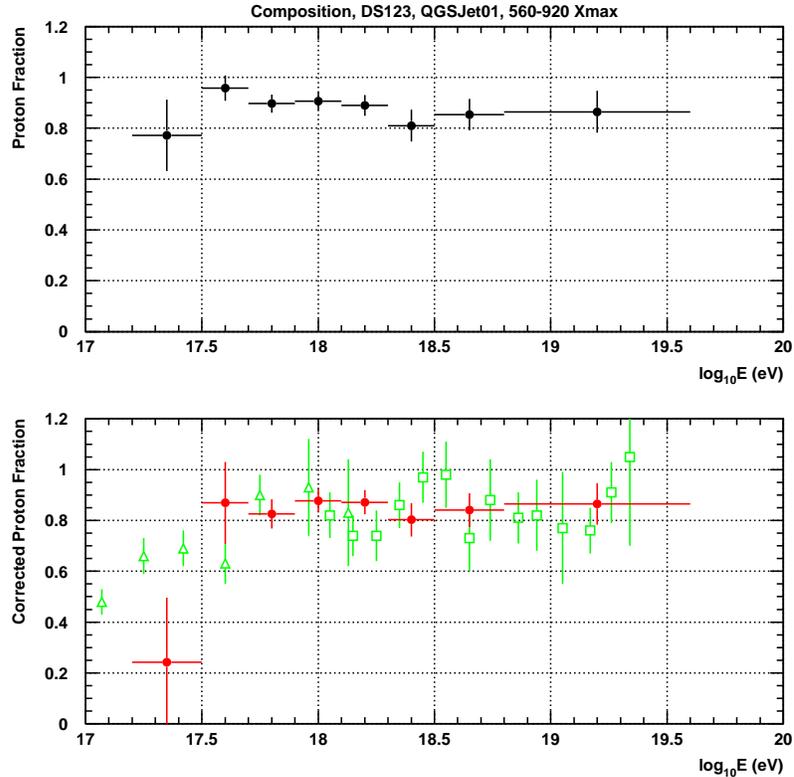}
    \caption{\label {compfit-res} The top plane shows the measured
      proton fraction in each energy bin.  The bottom plane shows the
      proton fraction after the acceptance correction.  The HiRes/MIA
      and Hires Stereo \Xmax\ measurements, interpreted as a proton
      fraction using the QGSJet01 expectations, are shown in green.}
  \end{center}
\end{figure}

The best fit proton fraction in each bin is plotted in the top half of
Figure~\ref{compfit-res}, along with the statistical uncertainty.
After adjusting for the different acceptances of protons and iron, one
finds the corrected composition as shown in the bottom of
Figure~\ref{compfit-res}.  The measurement indicates that the
composition is dominated by protons above an energy of $\log_{10}E =
17.6$.  The composition lower energies, but there are large
uncertainties.

\begin{figure}[h]
  \begin{center}
    \includegraphics*[width=0.75\textwidth,angle=0,clip]
                                    {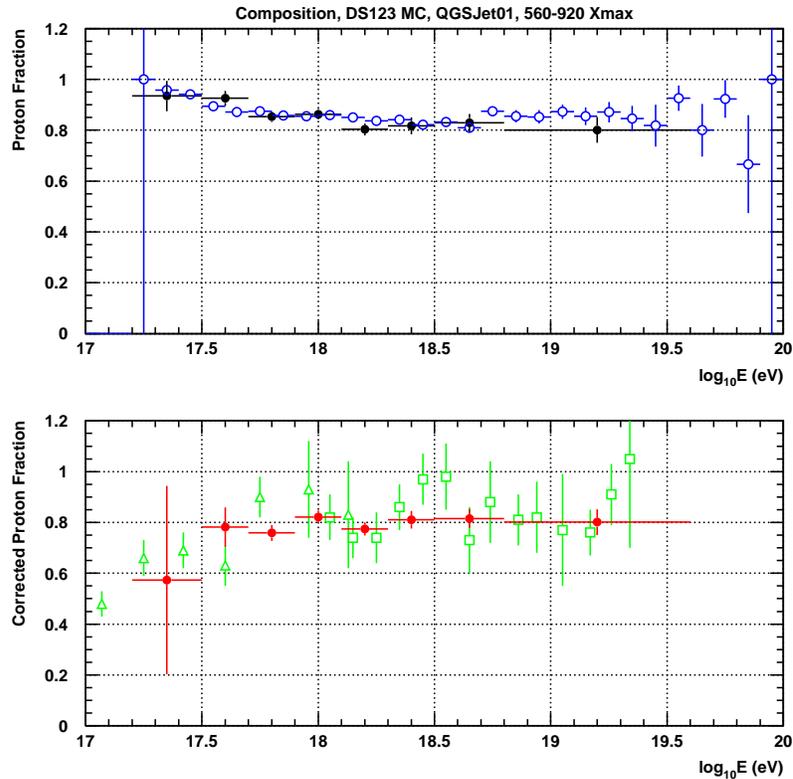}
    \caption{\label {compfit-res-mc} The top plane shows the measured
      proton fraction of a MC sample (filled black points), along with
      the exact fraction (open blue points).  The bottom plane shows
      the measured proton fraction of the MC sample after the
      acceptance correction.  The HiRes/MIA and Hires Stereo proton
      fractions, which were used as an input to the MC, are shown in
      green.}
  \end{center}
\end{figure}

As a cross check of our procedure, we have performed the same
analysis using the mixed MC.  The results, before and after the
acceptance correction, are shown in Figure~\ref{compfit-res-mc}.  In
the MC case, we can know the particle type of each shower that passes
the analysis cuts.  This is used to calculate the exact proton
fraction, which is shown as the blue points in the top plane of
Figure~\ref{compfit-res-mc}.  These points agree well with the proton
fractions obtained from fitting.  Likewise, the corrected MC
composition matches very well the HiRes/MIA and HiRes Stereo proton
fractions which were used as inputs.

\section{Conclusion}

We have made a preliminary measurement of the UHECR composition using
the HiRes-II detector.  This measurement indicates a light composition
above $\log_{10}E = 17.6$.  Measurements using an enlarged data set
and both QGSJet and Sibyll shower simulations will be presented in
Pune.

This work is supported by US NSF grants PHY-9321949, PHY-9322298,
PHY-9904048, PHY-9974537, PHY-0098826, PHY-0140688, PHY-0245428,
PHY-0305516, PHY-0307098, and by the DOE grant FG03-92ER40732. We
gratefully acknowledge the contributions from the technical staffs of
our home institutions. The cooperation of Colonels E.~Fischer and
G.~Harter, the US Army, and the Dugway Proving Ground staff is greatly
appreciated.

\end{document}